
\magnification=1200
\nopagenumbers
\def\truepageno{\footline={\hss\tenrm\folio\hss}}
\def\ni{\noindent}
\def\laq{\raise 0.4ex\hbox{$<$}\kern -0.8em\lower 0.62 ex\hbox{$\sim$}}
\def\gaq{\raise 0.4ex\hbox{$>$}\kern -0.7em\lower 0.62 ex\hbox{$\sim$}}
\baselineskip 15.7pt
\vsize=24.5truecm
\hsize=17.0truecm
\rightline{CERN-TH/95-36}
\rightline{DFTT 20/95}

\vskip 0.9cm
\centerline{{\bf String Theory and Cosmology}\footnote{$\diamond$}{This
essay was selected for an Honorable Mention by the Gravity Research
Foundation, 1995.}}
\vskip 1.0cm
\centerline{M. C. Bento\footnote{$^{*}$}{On leave of absence from
Departamento de F\' \i sica, Instituto Superior T\'ecnico, Av. Rovisco Pais,
1096 Lisboa Codex, Portugal.}}
\centerline{\it CERN, Theory Division}
\centerline{\it CH-1211  Geneva 23}
\centerline{\it Switzerland}
\vskip 0.2cm
\centerline{and}
\vskip 0.2cm
\centerline{O. Bertolami$^*$\footnote{$^\dagger$}{Also at Theory Division,
 CERN.}}
\centerline{INFN-\it Sezione Torino}
\centerline{\it Via Pietro Giuria 1}
\centerline{\it I-10125  Turin, Italy}

\vskip 1.5cm

\centerline{ABSTRACT}

\vskip 6mm

We discuss the main cosmological implications of considering string-loop
effects and a potential for the dilaton in the lowest
order string effective action.
Our framework is based on the effective model arising from regarding
homogeneous and isotropic dilaton, metric and Yang-Mills field configurations.
The issues of inflation, entropy crisis and the Polonyi problem as
well as the problem of the cosmological constant are discussed.
\vskip 1.0cm

\vfill
\eject
\truepageno\pageno=1

String theory, as a candidate to describe quantum gravity effects
and unify all
forces of nature is expected to account for particle interactions
at energies below the Planck scale and give rise to a consistent
cosmological scenario. For achieving these aims, string theory should lead
to the standard cosmological scenario and free this
model from its weaknesses, such as the existence of the cosmological
singularity and  initial condition problems. It is hence expected
that string low-energy models
naturally generate inflation, which is usually achieved through
a scalar field endowed with a potential. Furthermore, string models
should be free from the Polonyi problem associated with scalar fields
that couple only gravitationally as well as providing a mechanism
for the vanishing of the cosmological constant.

If, on the  one hand, the singularity problem seems to
admit a solution in the quantum cosmological framework, i.e. via
the vanishing of the wave function
for singular metric configurations [1], and a
radiation dominated era is shown to emerge from a string cosmological
scenario [2], the other difficulties, on the other hand,  do not have yet
a satisfactory explanation.
Indeed, none of the models proposed so far to break supersymmetry and
provide the dilaton with a potential, based on the
condensation of gauginos, seem to be
suitable for inflation [3]; moreover, no workable mechanism  has yet
been put forward to explain the vanishing of the cosmological constant [4].
In what regards the Polonyi problem, however,  some solutions
have been presented.
For instance, one can mention the  suggestion that string-loop effects
can satisfactorily drive the dynamics of
a massless dilaton [5] and the
possibility that the dilaton energy can be efficiently
transferred into coherent classical oscillations of Yang-Mills
fields [6].

In this essay, we adopt the point of view that supersymmetry is broken
by a mechanism other than gaugino condensation and, while not
specifying any such mechanism, we assume that a mass term for the
dilaton is thus generated and that this term dominates the dilaton
potential. We show that inflationary chaotic-type solutions can then
be obtained [6, 7]. We also discuss a possible solution to  the
Polonyi problem [6].
Finally, we present some ideas concerning the problem of the
cosmological constant.
We stress that the crucial ingredients of our discussion are the presence of a
potential for the dilaton and  string-loop effects. The
former is necessary not only to achieve inflation,
but also to break supersymmetry and to compensate the ensuing unbalanced
contribution of bosons and fermions to the vacuum energy. In what
concerns string-loop effects, it is shown that
their inclusion does not lead, on its own, to stable de Sitter
solutions [8], but they may be relevant for the Polonyi problem.

The four-dimensional string vacua emerging, for instance,
from heterotic string theories, correspond to N=1 non-minimal
supergravity and super Yang-Mills models [9]. The
bosonic action is, at lowest order in $\alpha^\prime$, given by:

$$ S_B=\int d^4x \sqrt{-g} \left\{ -{R\over 2 k^2} + 2 (\partial
\phi)^2 - B(\phi) Tr\left( F_{\mu\nu} F^{\mu\nu}\right) + 4
V(\phi) \right\} ,\eqno(1)$$

\ni
where $k^2=8\pi M_P^{-2}$, we have dropped the
antisymmetric tensor field $H_{\mu\nu\lambda}$ and introduced
a potential,  $V(\phi)$, for the dilaton as well as the universal function [5]

$$ B(\phi) = e^{- 2 k \phi} + c_0 + c_1 e^{2 k \phi} + c_2 e^{4 k \phi}
+ ...   .\eqno(2)$$

\ni
which expresses the fact that string-loop interactions
have an expansion in powers of $g_S \equiv e^{2 k \phi}$; the coefficients
$c_0, c_1, c_2$, ... are presently unknown.
The field strength $F_{\mu\nu}^a$ in (1) corresponds to the one of
a Yang-Mills theory with a gauge group G, which must be a subgroup of
$E_8\times E_8$.

As we are interested in a cosmological setting,  we
focus on  homogeneous and isotropic
field configurations on a spatially flat spacetime. The most general
metric is given in terms of the lapse function,
N(t), and the scale factor, a(t):

$$ds^2=-N^2(t) dt^2 + a^2(t) d\Omega^2_3 ~.\eqno(3)$$

We consider for simplicity an SO(3) gauge field;
our conclusions, however, will be qualitatively independent of
this choice. We then use the following homogeneous and
isotropic ansatz (up to a gauge
transformation) for the gauge potential [10, 11]:

$$ A_\mu (t) dx^\mu = \sum_{a,b,c=1}^{3} {\chi_0(t)\over 4} T_{a b}
\epsilon_{a c b } dx^c,\eqno(4)$$

\ni
$\chi_0(t) $ being an arbitrary function of time and $T_{ab}$ the generators
of SO(3).

We start by dimensionally reducing the system (1), allowing
only for homogeneous and isotropic field configurations.
This procedure allows one to treat the
contribution of the Yang-Mills fields on the same footing as the
remaining fields, as opposed to the usual treatment of radiation as a
fluid.

Introducing the ans\"atze (3) and (4) into the action (1) yields, after
integrating over $R^3$ and dividing by the infinite volume of its
orbits:

$$ S_{eff}=-\int_{t_1}^{t_2} dt \left\{ -{3 {\dot a}^2 a \over k^2 N}
+ {3 a\over N} B(\phi) \left[ {{\dot\chi_0}^2\over 2} - {N^2
\over a^2} {\chi_0^4\over 8}\right] + { 2 a^3\over N} {\dot \phi}^2 - 4
a^3 N V(\phi)\right\},\eqno(4)$$

\ni
where the dots denote time derivatives.
The equations of motion in
the N=1 gauge are the following:

$$\eqalignno{
             2 {\ddot a \over a} + H^2 + {k^2 \over 3} B(\phi)
\rho_{\chi_0} + 2 k^2  [ {\dot \phi}^2 - 2 V(\phi)] & = 0,  & (5) \cr
 \ddot \phi + 3 H {\dot \phi} - {1\over 4} {B^\prime(\phi)} \zeta_{\chi_o}
+ {\partial V(\phi) \over \partial \phi} & = 0, & (6)\cr
  {\ddot \chi_0} + [H+{B^\prime(\phi)\over B(\phi)}  \dot \phi] {\dot \chi_0} +
{\chi_0^3 \over 2 a^2} &
=0, & (7) \cr}$$

\ni
where the primes denote derivatives with respect to $\phi$, $H=\dot a / a$,
$\rho_{\chi_o}  = 3  \left[{{\dot \chi_0}^2\over 2 a^2} + {\chi_0^4 \over 8
a^4} \right]$ and
$\zeta_{\chi_0} =  3  \left[{{\dot \chi}_0^2\over 2 a^2} - {\chi_0^4 \over 8
a^4} \right]$.

Furthermore,  extremizing
action (4) with respect to N, yields the Friedmann equation:

$$H^2 = {k^2\over 3} \left[ 4 \rho_\phi + B(\phi)\rho_{\chi_o}\right]
\eqno(8)$$

\ni
with $\rho_\phi = {1\over 2} {\dot \phi}^2 + V(\phi)$.

Using our field treatment of
radiation, we now
discuss the way inflationary solutions can be obtained from the
dynamical system arising from eqs. (5)--(8). We follow the discussion of
ref. [7], where one obtains chaotic inflation
driven by the dilaton potential
$ V(\phi)= {1\over 2} m^2 (\phi -
\phi_0)^2$, with $ 10^{-8}M_P < m < 10^{-6}M_P$ and $\phi_0\sim M_P$, a
result which remains valid if we add a quartic term to the potential.
Furthermore, in this situation
$H \gg {B^\prime(\phi)\over B(\phi)} \dot \phi$, which
allows us to solve eq. (7) in the conformal time $d\xi \equiv a^{-1}(t) dt$,
the solution being given implicitly in terms of an elliptic function [12].
Moreover, we find that $\rho_{\chi_0}={C\over a^4(t)}$,
where $C$ is an integration constant. Substituting these results into
eqs. (5) and (6), we obtain, after introducing the dimensionless
variables $x\equiv {m\over \mu } (\phi - \phi_0),\ y\equiv {1\over \mu} \dot
\phi,\ z\equiv {1\over m} H$ and $\eta\equiv m t$,
where $\mu^2={3 m^2 / 2 k^2}$, the
following non-autonomous three-dimensional dynamical system

$$\eqalignno{ x_\eta & = y, &(9-a) \cr
              y_\eta & = - x - 3yz + C_1 {\zeta_{\chi_0}}(t)
                             {B^\prime(x)}, &(9-b)\cr
              z_\eta &= 2 x^2-y^2-2 z^2, &(9-c)\cr} $$

\ni
where the index $\eta$ denotes derivative with respect to $\eta$ and
$C_1={k^2 / 6 m^2}$.  The phase space of the system is the
region in $R^3$ characterized  by the constraint equation (8):

$$z^2 -x^2 -y^2= {k^2\over 3m^2 }  {C\over a^4}
B(x)\ .\eqno(10)$$

As argued in refs. [7, 11], during inflation, terms proportional to
$a^{-4}(t)$ in (10) and  (9-b), where $\zeta_{\chi_0}\sim a^{-4}$,
become much smaller than the remaining ones. Clearly this does
not hold if $x\ll 0$ and $x\gg 0$,  hence we disregard these regions
of phase space. Dropping
these terms, the resulting dynamical system has
in the finite region of variation of
$x,\ y,\ z$, a critical point at  the origin, which
for expanding models ($z>0$), is an asymptotically stable focus.
Hence, inflationary solutions do exist and inflation  with more than 65
e-foldings requires that the initial value of the $\phi$ field is such
that $\phi_{i}\gaq 4 M_P$ [7, 11].

Let us now consider the entropy crisis and Polonyi problems.
The former  difficulty concerns the dilution of the baryon
asymmetry generated prior to $\phi$ conversion into radiation.
The entropy crisis
problem can be solved either by regenerating in string-inspired models
the baryon asymmetry after $\phi$ decay [13] or by considering
the Affleck-Dine mechanism to
generate an ${\cal O}(1)$ baryon asymmetry and then allowing for its dilution
via $\phi$ decay [14]. In
models where the dilaton mass is very small, such that its lifetime is
greater than the age of the Universe,
one may  encounter the Polonyi problem if
$\rho_\phi$ dominates the energy density of the Universe at present.
These problems afflict various N=1
supergravity [15, 16] as well as string models [17] and
dynamical supersymmetry breaking scenarios [18].

 In what concerns avoiding the
Polonyi problem, a necessary requirement is that, at the time when $\phi$
becomes non-relativistic, i.e. $H(t_{NR})=m$, the ratio of
its energy density to the one of radiation satisfies [16]

$$ \epsilon \equiv {\rho_\phi(t_{NR})\over
\rho_{\chi_0}(t_{NR})}\laq 10^{-8}.\eqno(11)$$

\ni
Our field treatment of radiation reveals an energy exchange mechanism
that may be relevant in this context. Working out eqs. (6)--(8),
one obtains the energy exchange equations:

$$\eqalignno{{\dot \rho}_{\phi} & = - 3 H  {\dot
\phi}^2 +  {1\over 4} {B^\prime(\phi)}  \zeta_{\chi_o}
    {\dot \phi}, & (12)\cr
     {\dot \rho}_{\chi_0} & = -4 H \rho_{\chi_0}  - 3
  {B^\prime(\phi)\over B(\phi)} {\dot \chi_0^2\over a^2} \dot \phi ~.
&(13)\cr} $$

\ni
The new feature in eqs. (12), (13) are the
terms proportional to $\dot \phi$. Clearly, these terms do
not play any role in the reheating process, which is due to $\phi$ decay and
conversion into radiation as it quickly oscillates around the minimum of
its potential.

Notice that the condition,  $\Gamma_\phi^{-1}\geq t_U \approx 10^{60}\
M_P^{-1}$, implies
from the dilaton decay width $\Gamma_\phi \simeq
{m^3\over M_P^2}$ [14] that
$m\leq 10^{-20} M_P$, which falls outside the mass interval for which
inflation takes place; thus, we have to assume that, in models
where this problem occurs,  some  field
other than the dilaton, eg. moduli, $E_6$ singlets or scalars
of the hidden $E_8$ sector,
will drive inflation and be responsible for reheating.
In the absence of the
terms proportional to $\dot \phi$ in eqs. (12), (13)
and until $\phi$ becomes non-relativistic, $\rho_\phi\simeq
\rho_{\chi_0} \simeq {1\over 2} m^2 \phi_\ast^2$, implying that
$\epsilon = {\cal O} (1)$. Hence,  any mechanism for
draining $\phi$ energy into radiation has to be fairly efficient
to avoid the  Polonyi problem.
However, assuming that after
inflation the term proportional to
$\dot \phi$ in eq. (12) can effectively drain the dilaton energy into radiation
over the period ($t_i,\ t_{NR}$), during which
$H\simeq {B^\prime(\phi)\over B(\phi)}
\dot \phi$ (cf. eq. (7)), and that  $\phi_\ast\simeq \phi(t_{RN})\approx M_P$,
$\zeta_{\chi_0}\sim a^{-4}$ and $a(t)=  a_R \left({t\over t_R}\right)^{1/2}$,
where the index R refers to  the inflaton decay time, it then follows [6]

$$t_i\simeq {1\over m M_P}{t_R\over a_R^2}~.\eqno(14)$$

\ni
For typical values of the relevant parameters, e.g. $t_i\simeq 10^{10}
M_P^{-1}$, $t_R\gaq 10^{30} M_P^{-1}$ and \break $ a_R\gaq 10^{30}
M_P^{-1}$, and we see that the dilaton mass is required to be exceedingly
small, \break $m\leq 10^{-40} M_P$.
It is clear that an effective transfer of the dilaton
energy into radiation can be achieved if the conditions described above can be
maintained over a sufficiently long time interval and actually via
terms in $B(\phi)$ with negative coefficients.
Other contributions to this process
would be present if we had chosen a larger gauge
group as, besides $\chi_0(t)$, a multiplet of scalar fields would
appear in the effective action leading to extra energy
exchange terms [10, 11].

Let us now turn to the problem of the cosmological constant. Our mechanism
is inspired on the Atkin-Lehner symmetry
known to hold at 1-loop order. We start adding to action (1) the
contribution of bosons and fermions to obtain $S_T$. We then assume that
fermionic fields can be redefined such that they do not
interact with the dilaton [5] whereas
bosons do interact with  the dilaton via the universal $B(\phi)$ function. This
universality is related with the string S-duality
which implies that $B(\phi) = B(-\phi)$. Furthermore, we shall assume that
the contributions to the vacuum energy of bosons and fermions  depend on the
value of the dilaton field. Hence, at the minimum $\phi = \phi_0$:

$${\delta{S_{T}}\over \delta\phi} = \sqrt{-g}~
[{B^\prime(\phi_0)} V_B(\phi_0) + V^\prime(\phi_0)] = 0~,\eqno(15)$$
\ni
and, for the trace of the energy-momentum tensor, we have:

$$\Lambda(\phi_0) \equiv  \sqrt{-g}~[B(\phi_0) V_B(\phi_0) + V_F(\phi_0)
+ V(\phi_0)]~.\eqno(16)$$

Vanishing of  $\Lambda(\phi_0)$, for the value  $\phi = \phi_0$ that
satisfies eq. (15), implies that spacetime is flat and that
the cosmological constant vanishes. However, as
discussed in ref. [4], adjusting mechanisms that aim to dynamically solve the
cosmological constant problem are unable to satisfy both (15) and (16)
since one cannot usually preserve symmetries and achieve equilibrium
simultaneously. However, if one assumes that $\Lambda(\phi) = \Lambda(-\phi)$
and that at the ground-state
potentials do not respect this duality, i.e. $V(-\phi) = - V(\phi)$,
then one gets a vanishing
cosmological constant due to S-duality.
 Notice here  the
``awareness'' that boson and fermion vacuum contributions must have
of the value of the dilaton field at its minimum indicating that, in order
to fully implement string symmetries into the corresponding field theory,
non-local effects must be introduced. We stress that string theory
is rich in discrete and duality-type symmetries and it would not be
at all surprising that they  would play a role in solving
the cosmological constant problem.

\vfill
\eject
\vskip 2.0cm
\centerline{\bf References}
\vskip 0.8cm

\ni
\item{[1]} M.C. Bento and O. Bertolami, ``Scale Factor Duality: A
Quantum Cosmological Approach'', Preprint CERN-TH.7488/94, DFTT 43/94.

\ni
\item{[2]} A.A. Tseytlin and C. Vafa, Nucl. Phys. B372 (1992) 443.

\ni
\item{[3]}  P. Binetruy and M.K. Gaillard, Phys. Rev. D34 (1986) 3069;

R. Brustein and P.J. Steinhardt, Phys. Lett. B302 (1993) 196.

\ni
\item{[4]} S. Weinberg, Rev. Mod. Phys. 61 (1989) 1.

\ni
\item{[5]} T. Damour and A.M. Polyakov,  Nucl. Phys. B423 (1994) 532.

\ni
\item{[6]} M.C. Bento and O. Bertolami, Phys. Lett. B336 (1994) 6.

\ni
\item{[7]} M.C. Bento, O. Bertolami and P.M. S\'a, Phys. Lett. B262
(1991) 11.

\ni
\item{[8]} M.C. Bento and O. Bertolami, ``Cosmological Solutions of
Higher--Curvature String Efffective Theories with Dilatons'',
Preprint CERN-TH/95-63 , DFTT 19/95.

\ni
\item{[9]} E. Witten, Phys. Lett. B155 (1985) 151; Nucl. Phys. B268
(1986) 79.

\ni
\item{[10]} O. Bertolami, J.M. Mour\~ao,
R.F. Picken  and I.P. Volobujev, Int. J. Mod. Phys. A6 (1991) 4149.

\ni
\item{[11]} P.V. Moniz, J.M. Mour\~ao and P.M. S\'a, Class. Quantum
Gravity 10 (1993) 517.

\ni
\item{[12]} M.C. Bento, O. Bertolami, P.V. Moniz, J.M. Mour\~ao and
P.M. S\'a, Class. Quantum Gravity 10 (1993) 285.

\ni
\item{[13]} K. Yamamoto, Phys. Lett. B168 (1986) 341;

O. Bertolami and G.G. Ross, Phys. Lett. B183 (1987) 163.

\ni
\item{[14]} M.C. Bento, O. Bertolami and P.M. S\'a, Mod. Phys. Lett.
A7 (1992) 911.

\ni
\item{[15]} G.D. Coughlan, W. Fischler, E.W. Kolb, S. Raby and G.G.
Ross, Phys. Lett. B131 (1983) 59;

G.D. Coughlan, R. Holman, P. Ramond and G.G. Ross, Phys.
Lett. B140 (1984) 44;

A.S. Goncharov, A.D. Linde and M.I. Vysotsky, Phys. Lett. B147 (1984) 279;

O. Bertolami and G.G. Ross, Phys. Lett. B171 (1986) 46.

\ni
\item{[16]} O. Bertolami, Phys. Lett. B209 (1988) 277.

\ni
\item{[17]} B. de Carlos, J.A. Casas, F. Quevedo and E. Roulet, Phys. Lett.
B318
(1993) 447.

\ni
\item{[19]} T. Banks, D.B. Kaplan and A.E. Nelson, Phys. Rev. D49 (1994) 779.

\bye